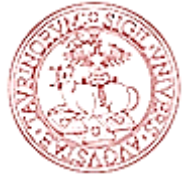

University of Turin

*Department of Economics and Statistics «Cognetti de Martiis»*

# CLIMATE EVENTS AND INSURANCE DEMAND

The effect of potentially catastrophic events

on insurance demand in Italy

Alessandro Chieppa

Andrea Ricca

Gianluca Rosso

Research coordinator: Chiara Daniela Pronzato

Collaboration: Isabella Pecetto

# ABSTRACT

Climate extreme events are constantly increasing. What's the effect of these potentially catastrophic events on insurance demand in Italy, with particular reference to the economic activities?

Extreme precipitation events over most of the mid-latitude land masses and over wet tropical regions will very likely become more intense and more frequent by the end of this century, as global mean surface temperature increases.

If we look to Italy, examination of the precipitation time series shows a sensitive and highly significant decrease in the total number of precipitation events in Italy (average of 12% from 1880 to the present), with a trend of events intense dissimilar as regards to low and high intensity, with a decline of firsts and an increase of seconds. The risk related to hydrological natural disasters is in Italy one of the most important problem for both damage and number of victims.

How evolves the ability to pay for damages, with a view to safeguarding work and economic activities, and employment protection?

GLOBAL WARMING.

Climate extreme events are constantly increasing.
What's the effect of these potentially catastrophic events on insurance demand in Italy, with particular reference to the economic activities?
We refers to the IPCC Intergovernmental Panel on Climate Change (Climate Change 2013, WG1 Working Group I, 5th Assessment Report).
Warming of the climate system is unequivocal, and since the 1950s, many of the observed changes are unprecedented over decades to millennia. The atmosphere and ocean have warmed, the amounts of snow and ice have diminished, sea level has risen, and the concentrations of greenhouse gases have increased.
Each of the last three decades has been successively warmer at the Earth's surface than any preceding decade since 1850. In the Northern Hemisphere, 1983–2012 was *likely* the warmest 30-year period of the last 1400 years (*medium confidence*).

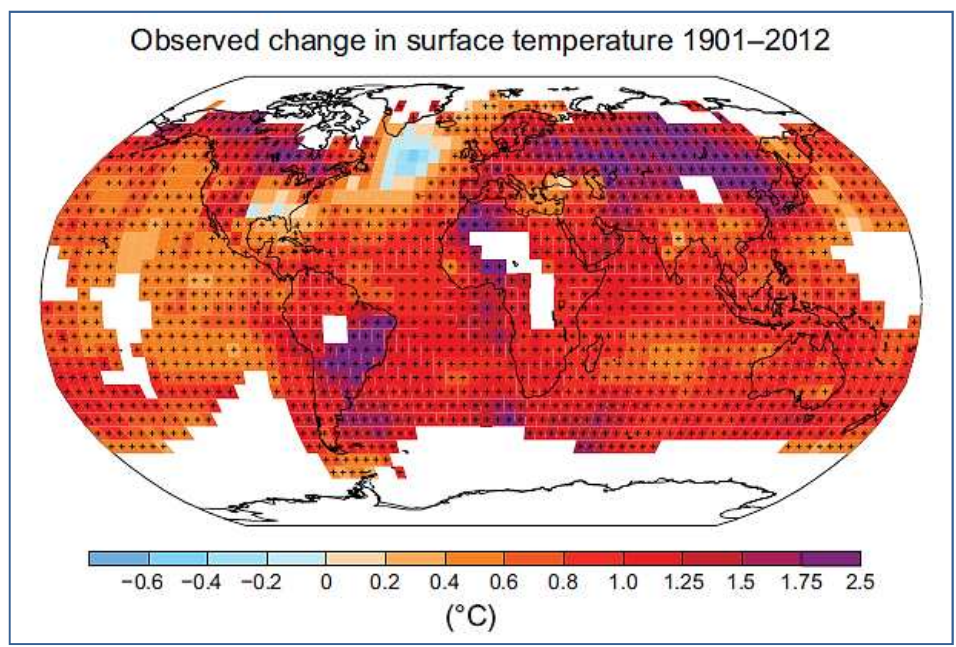

*Fig. 1. Source: IPCC Intergovernmental Panel on Climate Change - Climate Change 2013, WG1 Working Group I, 5th Assessment Report.*

Human influence has been detected in warming of the atmosphere and the ocean, in changes in the global water cycle, in reductions in snow and ice, in global mean sea level rise, and in changes in some climate extremes. This evidence for human influence has grown since AR4. It is extremely likely that human influence has been the dominant cause of the observed warming since the mid-20th century.

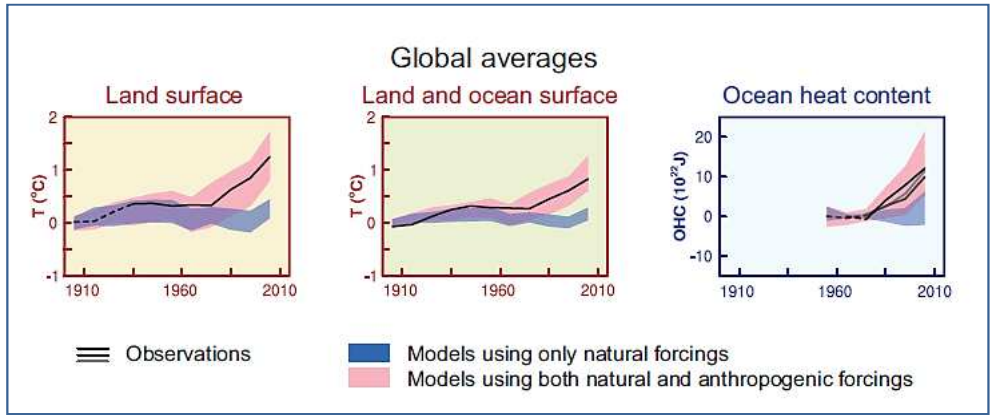

*Fig. 2. Source: IPCC Intergovernmental Panel on Climate Change - Climate Change 2013, WG1 Working Group I, 5th Assessment Report.*

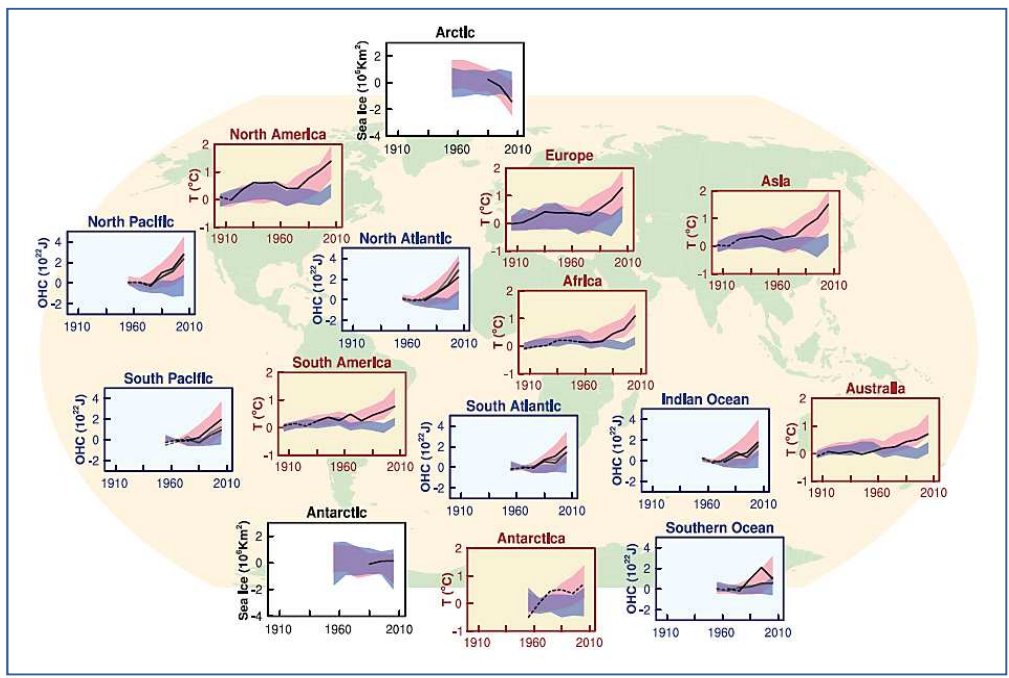

*Fig. 3. Source: IPCC Intergovernmental Panel on Climate Change - Climate Change 2013, WG1 Working Group I, 5th Assessment Report.*

Global surface temperature change for the end of the 21st century is likely to exceed 1.5°C relative to 1850 to 1900. Warming will continue beyond 2100. Warming will continue to exhibit interannual-to-decadal variability and will not be regionally uniform.

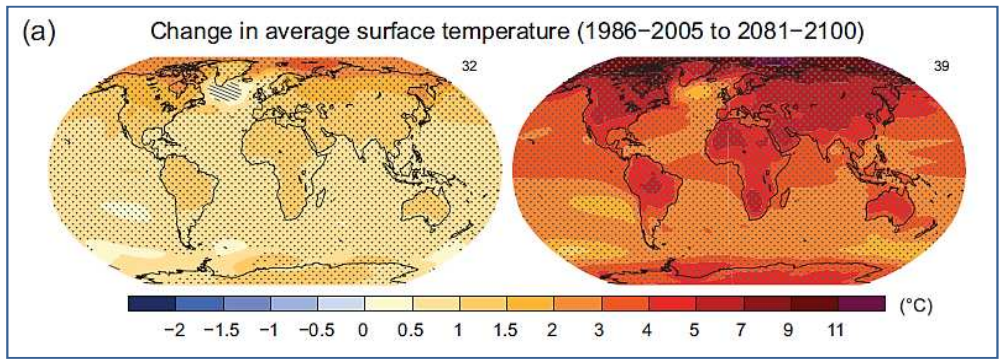

*Fig. 4. Source: IPCC Intergovernmental Panel on Climate Change - Climate Change 2013, WG1 Working Group I, 5th Assessment Report.*

Changes in the global water cycle in response to the warming over the 21st century will not be uniform. The contrast in precipitation between wet and dry regions and between wet and dry seasons will increase, although there may be regional exceptions.

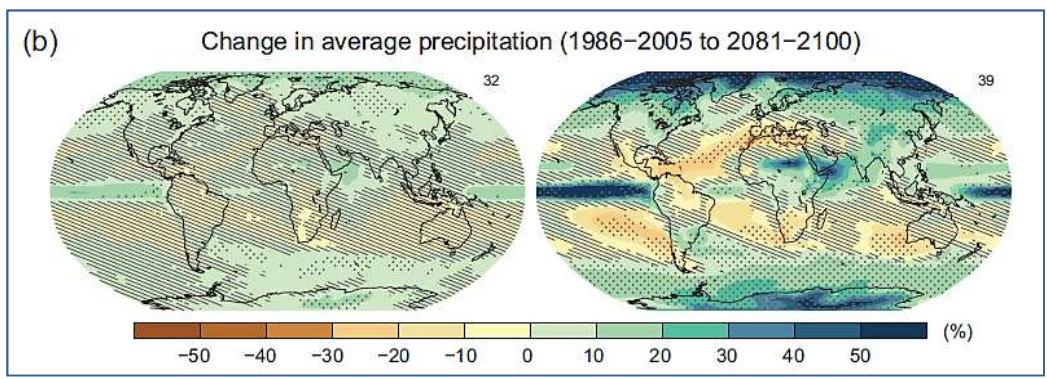

*Fig. 5. Source: IPCC Intergovernmental Panel on Climate Change - Climate Change 2013, WG1 Working Group I, 5th Assessment Report.*

Extreme precipitation events over most of the mid-latitude land masses and over wet tropical regions will very likely become more intense and more frequent by the end of this century, as global mean surface temperature increases.

EXTREME EVENTS.

Still refering to the IPCC Intergovernmental Panel on Climate Change (Climate Change 2013, WG1 Working Group I, 5th Assessment Report).

The impacts of climate extremes and the potential for disasters result from the climate extremes themselves and from the exposure and vulnerability of human and natural systems. Observed changes in climate extremes reflect the influence of anthropogenic climate change in addition to natural climate variability, with changes in exposure and vulnerability influenced by both climatic and non-climatic factors.

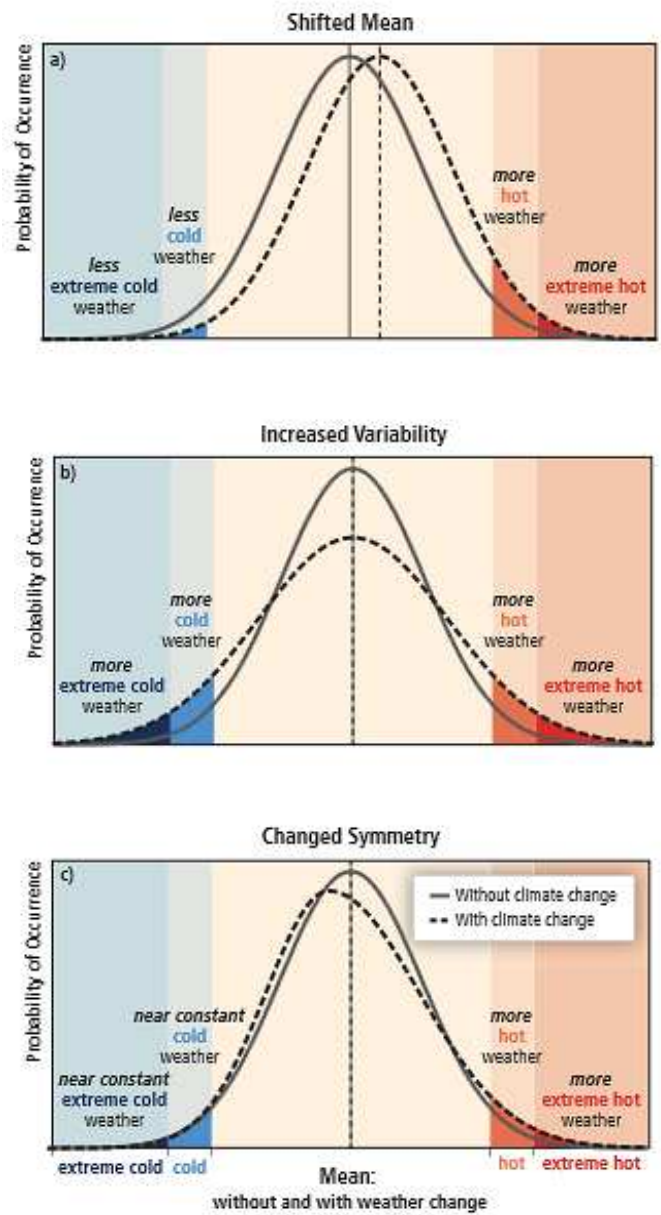

*Figg. 6-7-8. Source: IPCC Intergovernmental Panel on Climate Change - Climate Change 2013, WG1 Working Group I, 5th Assessment Report.*

Economic losses from weather- and climate-related disasters have increased, but with large spatial and interannual variability (high confidence, based on high agreement, medium evidence). Global weather- and climate-related disaster losses reported over the last few decades reflect mainly monetized direct damages to assets, and are unequally distributed. Loss estimates are lowerbound estimates because many impacts, such as loss of human lives, cultural heritage, and ecosystem services, are difficult to value and monetize, and thus they are poorly reflected in estimates of losses. Impacts on the informal or undocumented economy as well as indirect economic effects can be very important in some areas and sectors, but are generally not counted in reported estimates of losses.

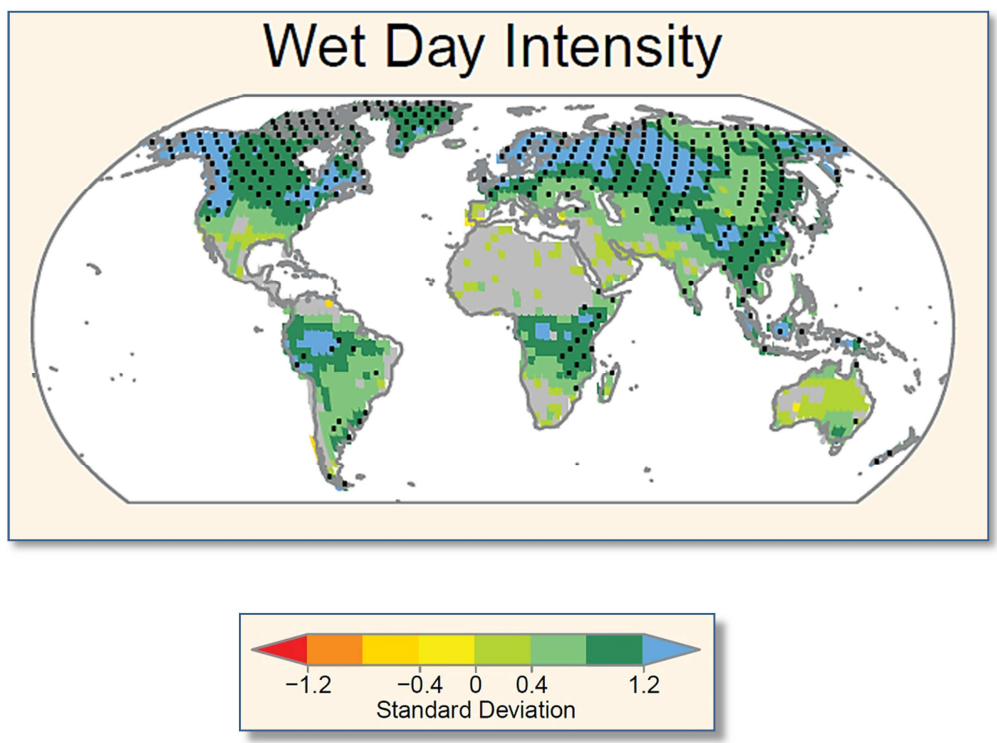

*Fig. 9. Source: IPCC Intergovernmental Panel on Climate Change - Climate Change 2013, WG1 Working Group I, 5th Assessment Report.*

"If we look to Italy, examination of the precipitation time series shows a sensitive and highly significant decrease in the total number of precipitation events in Italy (average of 12% from 1880 to the present), with a trend of events intense dissimilar as regards to low and high intensity, with a decline of firsts and an increase of seconds.

No wonder if in a world more "hot", where precisely the "Gaussian" temperature has aready moved towards higher values may occur more extreme events, both related to the thermal field and rainfall, and which it cause more heat waves, drought and more intense perturbations (it's already much more controversial if they can be even more numerous.)" (*Carlo Cacciamani, Arpa Emilia-Romagna*).

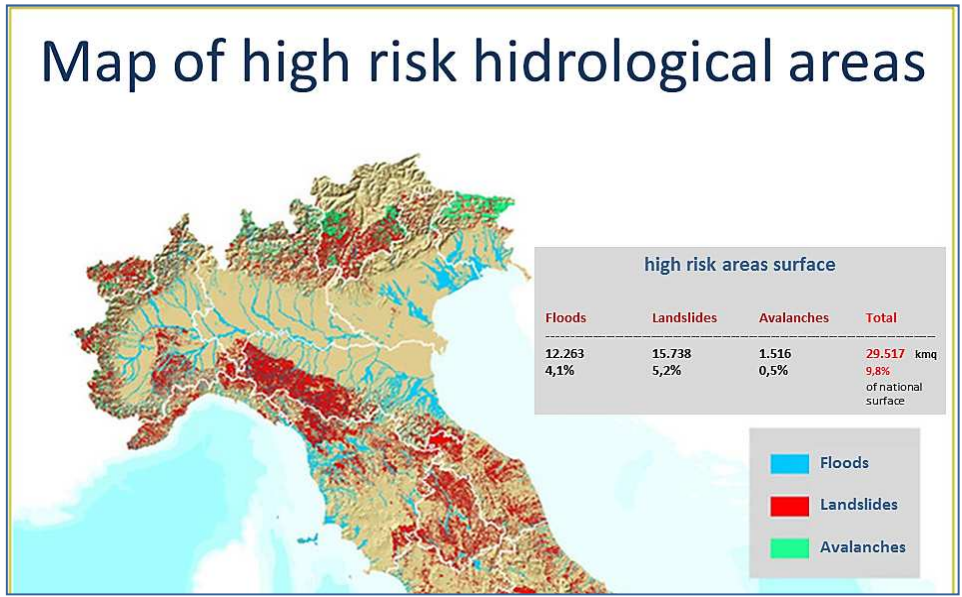

*Fig. 10.*

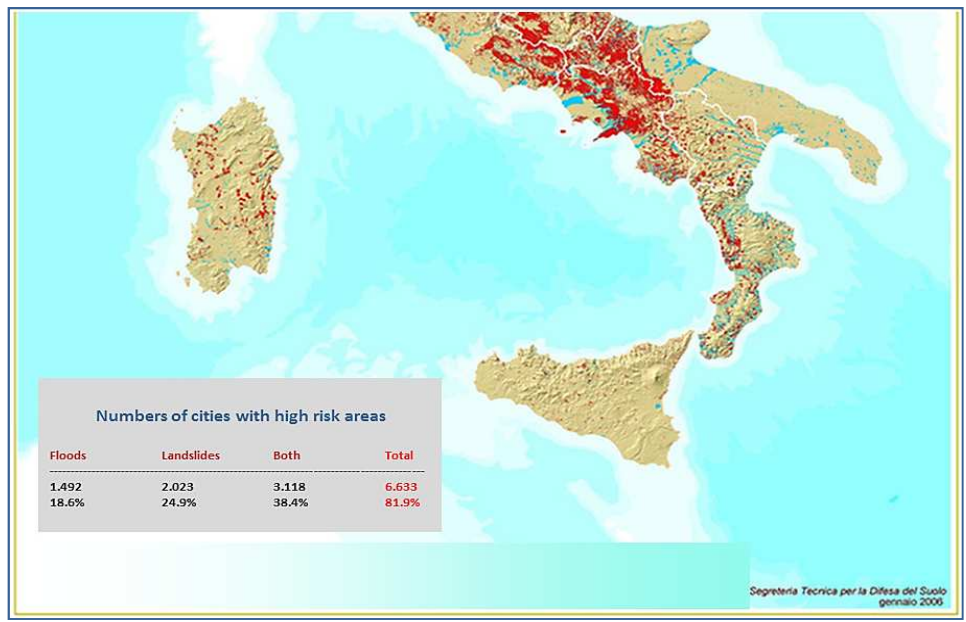

*Fig. 11.*

"The risk related to hydrological natural disasters is in Italy one of the most important problem for both damage and number of victims. The increasing impact of hydrological disasters in the area, since the war, it should first be attributed to the changed scenarios that favored occupation and exploitation of naturalized areas and, marginally, to weather and climate changes. In practice, most of the damage resulting from the hydrological instability are mainly determined by human behavior and by practical development models, rather than by an presumed increase in natural hazard in the area." (*Willer Bordon, Ministro dell'Ambiente*).

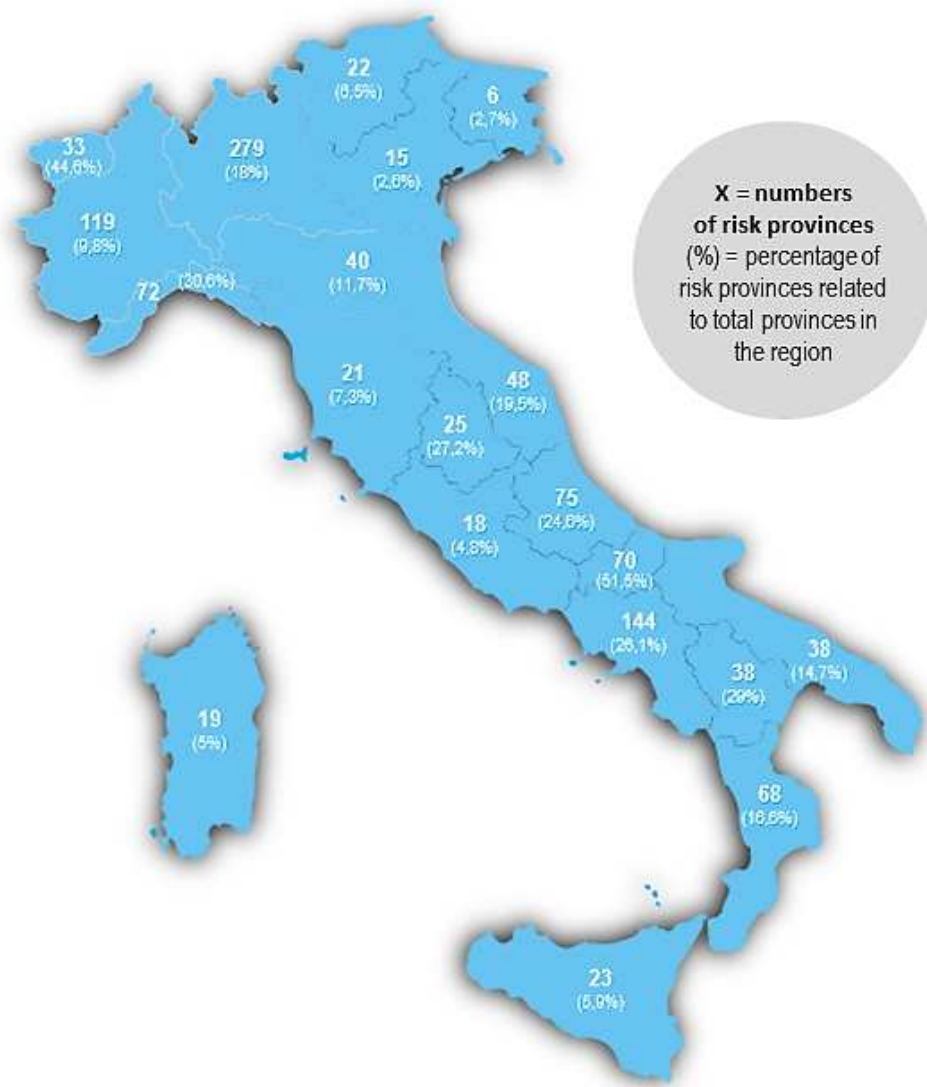

*Fig. 12.*

RESEARCH.

As said, we know that extreme weather events are increasing constantly, not only in frequency, but also in intensity and their destructive power. And Italy by the hydrogeological point of view is not in an optimal condition.

How evolves the ability to pay for damages, with a view to safeguarding work and economic activities, and employment protection?

The current Italian system is based on ad hoc approved refunding after a catastrophic event.

Historically, the public compensation did not cover the total amount of damage, but a percentage ranging from 50% to 80%.

The current system does not provide incentives for prevention and mitigation of risk, and especially not aside reserves to cover for events, reserves that, if invested, would compete to mitigate the cost.

| EVENTS | COVERAGE DIFFUSION | OFFERS AVAILABILITY |
|---|---|---|
| **CLIMATE EVENTS** | **High** (>30%) | **High** (always offered without selection) |
| **EARTHQUAKES** | **Medium** (5% to 30%) | **Medium** (offered with low underwriting selection) |
| **FLOODS** | **Medium** (5% to 30%) | **Medium** (offered with low underwriting selection) |
| **LANDSLIDES** | **Negligible** (<1%) | **Negligible** (never offerd) |
| **ERUPTIONS** | **Negligible** (<1%) | **Low** (offered with strong selection) |
| **SNOW** | **Medium** (5% to 30%) | **Medium** (offered with low underwriting selection) |

*Tab. 1. Source: LO STATO DELL'ARTE SULL' ASSICURAZIONE DEI RISCHI CATASTROFALI, R. Manzato, ANIA.*

From the surveys Perils for the corporate insurance market segment, there is a significant exposure to catastrophic risks (flood and earthquake), equal to, in 2013, about 300 billion euro (sums insured).

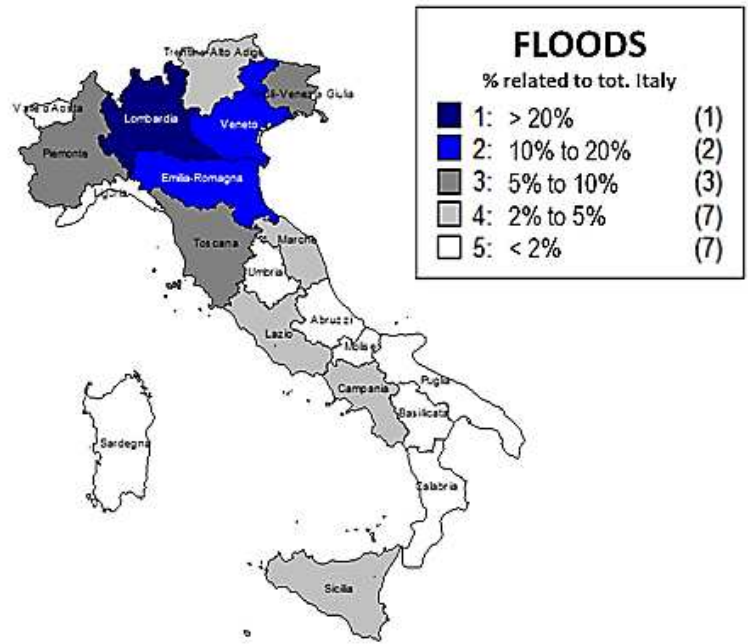

*Fig. 13. Source: LO STATO DELL'ARTE SULL' ASSICURAZIONE DEI RISCHI CATASTROFALI, R. Manzato, ANIA.*

These exposures are concentrated especially in the most industrialized regions, i.e. Lombardy, Emilia Romagna and Veneto.

DATASETS.

Reserach used two different datasets:
- the first one relative to italian climate data from last eleven years (2003-2013);
- the second one relative to insurance data from a primary italian Insurance Company.

Climate dataset.
For each month from Jan 2003 to Dec 2013, the database contains:
- Region;
- province;
- Average Temperature °C;
- Total average rainfalls (mm) peak excluded;
- Average wind km/h;
- Max wind km/h;
- Day or days in the month where occured the catastrofic event (for us is flood);

- Number of dead (direct or indirect causes);
- Total days event duration;
- Affected provinces;
- Event rainfalls mm;
- Monetary quantification of the damage.

List of extreme weather events has been provided by SMI - Italian Meteorological Society.

Were detected general climate events in 88 provinces compared to 110 provinces in Italy, but all extreme events occurred are listed in the database.

Climate dataset summarize shows 34 months in which occured at least one extreme event in period 2003-2013.

```
. sum

    Variable |     Obs        Mean    Std. Dev.       Min        Max
-------------+--------------------------------------------------------
     regione |       0
   provincia |       0
        anno |   11062    2007.973    3.135485       2003       2013
        mese |   11062    6.504701    3.452118          1         12
      tmediac|   11062    15.44323    7.386522      -1.45      32.03
-------------+--------------------------------------------------------
 pioggiamed~h|   11062    83.93602    64.48151          0     1176.7
 ventomedio~h|   11062     8.92498    4.098252          0      30.85
  ventomaxkmh|   11062    39.79407    19.62325          0        100
 dateeventi~a|       0
 numeroeven~e|      34    1.029412    .1714986          1          2
-------------+--------------------------------------------------------
  mortitotali|      34    3.823529    6.762063          0         36
 duratatota~i|      34    1.823529    2.492055          1         15
   cittcolpite|      0
 precipextr~m|      34    291.7471    131.8975        105        542
 stimadanni~o|       0
-------------+--------------------------------------------------------
        danni|       0
```

Tab. 2.

Months without extreme events was 11,028. In 34 months with extreme events, only one had two events, and the others had a single event.

```
. tab evento

      evento |      Freq.     Percent        Cum.
-------------+-----------------------------------
           0 |     11,028       99.69       99.69
           1 |         34        0.31      100.00
-------------+-----------------------------------
       Total |     11,062      100.00
```

Tab. 3.

```
. tab numero,m

numeroevent
iextranelme
         se     Freq.     Percent        Cum.

          1        33        0.30        0.30
          2         1        0.01        0.31
          .    11,028       99.69      100.00

      Total    11,062      100.00
```

*Tab. 4.*

Extreme events without dead was 9 (11,037 «events zero dead» - 11,028 «months without extreme events»). Events with dead are distributed as shown in tabulation. Only one extreme event had a particularly high mortality (36 dead).

```
. tab1 morti precipitazioni

-> tabulation of morti

      morti     Freq.     Percent        Cum.

          0    11,037       99.77       99.77
          1         5        0.05       99.82
          2         6        0.05       99.87
          3         3        0.03       99.90
          4         4        0.04       99.94
          5         2        0.02       99.95
          6         2        0.02       99.97
         12         1        0.01       99.98
         18         1        0.01       99.99
         36         1        0.01      100.00

      Total    11,062      100.00
```

*Tab. 5.*

```
-> tabulation of precipitazioni

precipitazi
        oni     Freq.     Percent        Cum.

          0    11,028       99.69       99.69
        105         1        0.01       99.70
        115         1        0.01       99.71
        150         1        0.01       99.72
        154         1        0.01       99.73
        163         1        0.01       99.74
        190         1        0.01       99.75
      195.2         2        0.02       99.76
        198         1        0.01       99.77
        200         4        0.04       99.81
        203         1        0.01       99.82
        205         1        0.01       99.83
        213         1        0.01       99.84
        225         1        0.01       99.85
        230         1        0.01       99.86
        250         1        0.01       99.86
        265         1        0.01       99.87
        312         1        0.01       99.88
        372         1        0.01       99.89
        381         1        0.01       99.90
        385         1        0.01       99.91
        396         1        0.01       99.92
        400         3        0.03       99.95
        414         1        0.01       99.95
        485         1        0.01       99.96
        517         2        0.02       99.98
        542         2        0.02      100.00

      Total    11,062      100.00
```

*Tab. 6.*

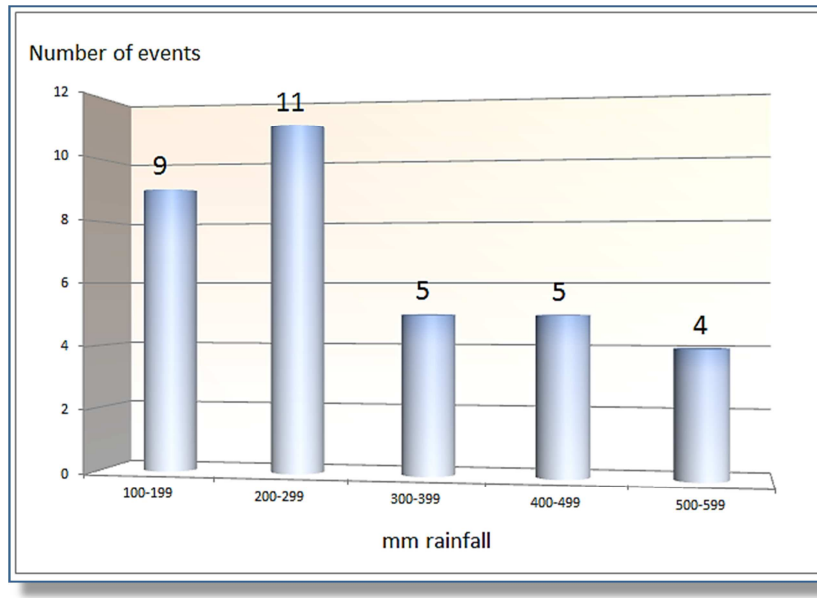

*Fig. 14.*

Insurance dataset.

A record for each issued policy in period 2003-2013.

Each record contains:

- Policy effect date;
- Policy issue date;
- Product type;
- Region of Contractor;
- province of Contractor.

Data are from a portion of database of an italian Casualty primary Insurance Company, and are absolutely anonimous.

Products usefull for the research are two: they are not «stand alone» but is present the additional peril «flood».

Into the anlysis they are indicated as:

- Class 1
- Class 2

| ramoufficiopol | Freq. | Percent | Cum. |
|---|---|---|---|
| (01) INCENDIO | 9,442 | 49.96 | 49.96 |
| (14) RISCHI TECNOLOGICI | 9,456 | 50.04 | 100.00 |
| Total | 18,898 | 100.00 | |

*Tab. 7.*

It was generated a single database merging the insurance database with the climate one.

Items present in both databases are 18,898: they are number of months in which we have both climate events and policies (i.e. we know presence in both datasets), related to each province and each kind of policy (class 1 and class 2).

```
. tab _m

    _merge |      Freq.     Percent        Cum.
-----------+-----------------------------------
         1 |      7,921       29.52       29.52
         2 |         11        0.04       29.56
         3 |     18,898       70.44      100.00
-----------+-----------------------------------
     Total |     26,830      100.00
```

*Tab. 8.*

```
. tab provincia

    provincia |      Freq.     Percent        Cum.
--------------+-----------------------------------
       ANCONA |        262        1.39        1.39
       AREZZO |        258        1.37        2.75
         BARI |        264        1.40        4.15
      BELLUNO |        250        1.32        5.47
      BERGAMO |        264        1.40        6.87
      BOLOGNA |        264        1.40        8.27
      BOLZANO |        196        1.04        9.30
      BRESCIA |        264        1.40       10.70
     BRINDISI |        246        1.30       12.00
     CAGLIARI |        262        1.39       13.39
   CAMPOBASSO |        228        1.21       14.59
      CATANIA |        261        1.38       15.98
    CATANZARO |        232        1.23       17.20
       CHIETI |        260        1.38       18.58
         COMO |        261        1.38       19.96
      COSENZA |        261        1.38       21.34
      CREMONA |        261        1.38       22.72
      CROTONE |        181        0.96       23.68
        CUNEO |        264        1.40       25.08
         ENNA |        169        0.89       25.97
      FERRARA |        257        1.36       27.33
      FIRENZE |        264        1.40       28.73
       FOGGIA |        253        1.34       30.07
    FROSINONE |        241        1.28       31.34
       GENOVA |        264        1.40       32.74
      GORIZIA |        148        0.78       33.52
```

*Tab. 9.*

## METHODS AND REGRESSIONS.

In the months following an extreme event, how much insurance are issued compared to how much it would be issued in the absence of the event?
Two different regressions were done, one for each class of insurance.

$$\text{Number\_policies}_{t,i} = \beta * \text{event12m}_{t,i} + \varepsilon_{i,t}$$
$$(\text{Numero\_polizze}_{t,i} = \beta * \text{evento12m}_{t,i} + \varepsilon_{i,t})$$

- Number_policies = new policies at month *t* in province *i*
- event12m = extreme event in last 12 months from *t* in province *i*
- ε = casual error

As a first step has been checked the effect that a flood event within twelve months, omitting other variables. The effect is significative only for the Class 2, while the Class 1 already at this early stage is not affected by the presence of extreme events.

Class 1

```
. reg numero_polizze  evento12m if( ramo1 == 1 & calendario >= 13)

      Source |       SS       df       MS              Number of obs =    8547
-------------+------------------------------           F(  1,  8545) =    0.58
       Model |  142.383618     1   142.383618          Prob > F      =  0.4479
    Residual |  2112155.5   8545   247.180281          R-squared     =  0.0001
-------------+------------------------------           Adj R-squared = -0.0000
       Total |  2112297.88  8546   247.168018          Root MSE      =  15.722

  numero_pol~e |    Coef.   Std. Err.      t    P>|t|     [95% Conf. Interval]
-------------+----------------------------------------------------------------
    evento12m | -.7419354   .977559    -0.76   0.448    -2.658187    1.174316
        _cons |  11.66703  .1727793    67.53   0.000     11.32834    12.00572
```

*Tab. 10.*

Class 2

```
. reg numero_polizze  evento12m if( ramo2 == 1 & calendario >= 13)

      Source |       SS       df       MS              Number of obs =    8594
-------------+------------------------------           F(  1,  8592) =    3.11
       Model |  587.69174     1   587.69174           Prob > F      =  0.0779
    Residual |  1623702.25  8592   188.978381          R-squared     =  0.0004
-------------+------------------------------           Adj R-squared =  0.0002
       Total |  1624289.94  8593   189.024781          Root MSE      =  13.747

  numero_pol~e |    Coef.   Std. Err.      t    P>|t|     [95% Conf. Interval]
-------------+----------------------------------------------------------------
    evento12m |  1.504482  .8531357    1.76   0.078    -.1678694    3.176832
        _cons |  10.84626  .1506565   71.99   0.000     10.55094    11.14159
```

*Tab. 11.*

By controlling the distribution of policy numbers in the months through the use of dummies drivers, the significance for the Class 1 continues to be null, while for the Class 2 is weakened because of the fact that the months with significance do not match with the months calendar in which normally occurring flood events.

Class 1

```
. reg numero_polizze  evento12m month* if( ramo1 == 1 & calendario >= 13)

      Source |       SS       df       MS              Number of obs =    8547
-------------+------------------------------           F( 12,  8534) =   24.59
       Model |  70607.4536     12   5883.95447         Prob > F      =  0.0000
    Residual |  2041690.43   8534   239.241907         R-squared     =  0.0334
-------------+------------------------------           Adj R-squared =  0.0321
       Total |  2112297.88   8546   247.168018         Root MSE      =  15.467

numero_pol~e |      Coef.   Std. Err.      t    P>|t|     [95% Conf. Interval]
-------------+----------------------------------------------------------------
    evento12m |  -.7532646   .9618112    -0.78   0.434    -2.638647    1.132118
       month1 |   -5.94962   .8112923    -7.33   0.000    -7.539949    -4.35929
       month2 |  -7.633893   .8141416    -9.38   0.000    -9.229807   -6.037978
       month3 |  -6.095613   .8129893    -7.50   0.000    -7.689269   -4.501958
       month4 |  -7.542789    .813853    -9.27   0.000    -9.138138    -5.94744
       month5 |  -7.059884   .8115697    -8.70   0.000    -8.650757   -5.469011
       month6 |  -4.218481    .807406    -5.22   0.000    -5.801192   -2.635769
       month7 |  -6.407586   .8127104    -7.88   0.000    -8.000695   -4.814477
       month8 |  -12.47554   .8315272   -15.00   0.000    -14.10554   -10.84555
       month9 |  -8.806514   .8158919   -10.79   0.000    -10.40586   -7.207168
      month10 |  -8.377686   .8147219   -10.28   0.000    -9.974738   -6.780634
      month11 |  -9.244385   .8188614   -11.29   0.000    -10.84955   -7.639218
        _cons |   18.58561    .572955    32.44   0.000     17.46248    19.70874
```

*Tab. 12.*

Class 2

```
. reg numero_polizze  evento12m month* if( ramo2 == 1 & calendario >= 13)

      Source |       SS       df       MS              Number of obs =    8594
-------------+------------------------------           F( 12,  8581) =    8.09
       Model |   18170.735     12   1514.22792         Prob > F      =  0.0000
    Residual |  1606119.21   8581   187.171566         R-squared     =  0.0112
-------------+------------------------------           Adj R-squared =  0.0098
       Total |  1624289.94   8593   189.024781         Root MSE      =  13.681

numero_pol~e |      Coef.   Std. Err.      t    P>|t|     [95% Conf. Interval]
-------------+----------------------------------------------------------------
    evento12m |    1.51091   .8491691    1.78   0.075     -.153666    3.175485
       month1 |  -.4063101   .7269103   -0.56   0.576    -1.831229    1.018609
       month2 |  -.3518765    .727157   -0.48   0.628    -1.777279    1.073526
       month3 |   .9732354   .7248868    1.34   0.179     -.447717    2.394188
       month4 |  -.0216314   .7266495   -0.03   0.976    -1.446039    1.402776
       month5 |   1.416699   .7234104    1.96   0.050    -.0013592    2.834757
       month6 |   1.823802   .7231654    2.52   0.012     .4062238     3.24138
       month7 |   2.231362    .725136    3.08   0.002     .8099207    3.652803
       month8 |  -3.668699   .7367475   -4.98   0.000    -5.112901   -2.224497
       month9 |   .4411226   .7243902    0.61   0.543    -.9788564    1.861102
      month10 |   .5604933   .7226681    0.78   0.438      -.85611    1.977097
      month11 |  -.5552617   .7212156   -0.77   0.441    -1.969018    .8584944
        _cons |   10.61949    .516069   20.58   0.000     9.607867    11.63111
```

*Tab. 13.*

FIXED EFFECTS.

Two different regressions were done, one for each class of insurance, with the presence of fixed effect *u* to analyze the variations within individual provinces, so as to separate the events of each province; were also considered dummy variables to control the effect of the passing of months and years.

**Number_policies$_{t,i}$ = β\*event12m$_{t,i}$ + recurrence_month$_{t,i}$ + year$_{t,i}$ + u$_i$+ ε$_{i,t}$**
(Numero_polizze$_{t,i}$ = β\*evento12m$_{t,i}$ + mese_di_ricorrenza$_{t,i}$ + anno$_{t,i}$ + u$_i$+ ε$_{i,t}$)

- Number_policies = new policies at month *t* in province *i*
- event12m = extreme event in last 12 months from *t* in province *i*
- U= fixed effect for province *i*
- Recurrence month = dummy driver with value 1 in month of issue
- year = dummy driver as control of number of policies during the years, in month *t* in province *i*
- ε = casual error

It should be pointed out that there is a negative effect of years due to the fact that it's important to take into account the effect generated by the economic recession with regard to the number of policies issued to be insured against floods.

Class 1

```
. xtreg numero_polizze evento12m month* year* if( ramo1 == 1 & calendario >= 13) , fe i(id)

Fixed-effects (within) regression               Number of obs      =      8547
Group variable (i): id                          Number of groups   =        77

R-sq:   within  = 0.2451                        Obs per group: min =        41
        between = 0.2223                                       avg =     111.0
        overall = 0.0908                                       max =       120

                                                F(21,8449)         =    130.65
corr(u_i, Xb)  = -0.0342                        Prob > F           =    0.0000

  numero_pol~e |     Coef.   Std. Err.      t    P>|t|     [95% Conf. Interval]
     evento12m |   .8194356   .6088896     1.35   0.178     -.374137    2.013008
        month1 |  -6.097217   .4753393   -12.83   0.000    -7.028999   -5.165436
        month2 |  -7.972837   .4771732   -16.71   0.000    -8.908214   -7.037461
        month3 |  -6.335782   .4764798   -13.30   0.000    -7.269799   -5.401765
        month4 |  -7.822425   .4768856   -16.40   0.000    -8.757238   -6.887613
        month5 |  -7.246241   .4755417   -15.24   0.000     -8.17842   -6.314063
        month6 |    -4.1576   .4731381    -8.79   0.000    -5.085067   -3.230134
        month7 |  -6.627047    .476261   -13.91   0.000    -7.560635   -5.693458
        month8 | -13.59466    .4875003   -27.89   0.000    -14.55028  -12.63904
        month9 |  -9.169998     .47815   -19.18   0.000    -10.10729   -8.232707
       month10 |  -8.683881   .4774495   -18.19   0.000    -9.619799   -7.747964
       month11 |  -9.732187   .4800404   -20.27   0.000    -10.67318   -8.791191
         year3 |  -.4136465    .429318    -0.96   0.335    -1.255215    .4279219
         year4 |   .3291208   .4297934     0.77   0.444    -.5133795    1.171621
         year5 |  -1.641549   .4304074    -3.81   0.000    -2.485253   -.7978452
         year6 |  -2.037514    .431087    -4.73   0.000     -2.88255   -1.192478
         year7 |  -4.052614   .4325043    -9.37   0.000    -4.900428    -3.2048
         year8 |  -5.755845   .4335356   -13.28   0.000    -6.605681   -4.906009
         year9 |  -8.589889   .4387981   -19.58   0.000    -9.450041   -7.729737
        year10 | -10.87751    .4442192   -24.49   0.000    -11.74829  -10.00674
        year11 | -11.22408    .4473713   -25.09   0.000    -12.10104  -10.34712
         _cons |   23.0528    .4416981    52.19   0.000     22.18696   23.91863
       sigma_u |  11.860596
       sigma_e |  9.0604979
           rho |   .63148555   (fraction of variance due to u_i)

F test that all u_i=0:     F( 76, 8449) =   196.61              Prob > F = 0.0000
```

*Tab. 14.*

Class 2

```
. xtreg numero_polizze  evento12m month* year* if( ramo2 == 1 & calendario >= 13), fe i(id)

Fixed-effects (within) regression               Number of obs      =        8594
Group variable (i):  id                         Number of groups   =          77

R-sq:  within  = 0.1346                         Obs per group: min =          21
       between = 0.0953                                        avg =       111.6
       overall = 0.0230                                        max =         120

                                                F(21,8496)         =       62.94
corr(u_i, Xb)  = -0.0126                        Prob > F           =      0.0000

------------------------------------------------------------------------------
numero_pol~e |      Coef.   Std. Err.      t    P>|t|     [95% Conf. Interval]
-------------+----------------------------------------------------------------
   evento12m |   .1631627   .3829879     0.43   0.670    -.5875868    .9139121
      month1 |  -.3149127   .3034863    -1.04   0.299    -.9098196    .2799943
      month2 |  -.2832221   .3035568    -0.93   0.351    -.8782673    .3118232
      month3 |   1.130582   .3026288     3.74   0.000     .5373564    1.723809
      month4 |   .0451799   .3032939     0.15   0.882    -.5493499    .6397096
      month5 |   1.652193   .3019626     5.47   0.000     1.060273    2.244113
      month6 |   2.069507   .3018621     6.86   0.000     1.477784     2.66123
      month7 |   2.385579   .3027172     7.88   0.000      1.79218    2.978978
      month8 |  -4.003326   .3075962   -13.01   0.000    -4.606289   -3.400362
      month9 |   .6108626   .3023697     2.02   0.043     .0181443    1.203581
     month10 |   .7946191   .3016608     2.63   0.008     .2032907    1.385948
     month11 |   -.216707    .301037    -0.72   0.472    -.8068128    .3733988
       year3 |  -.4046892   .2759851    -1.47   0.143    -.9456872    .1363087
       year4 |  -.5676074   .2755057    -2.06   0.039    -1.107666   -.0275492
       year5 |   .3234366   .2769845     1.17   0.243    -.2195203    .8663936
       year6 |   .8141031   .2753354     2.96   0.003     .2743787    1.353827
       year7 |   .5555603   .2745892     2.02   0.043     .0172988    1.093822
       year8 |   1.982738   .2746391     7.22   0.000     1.444378    2.521097
       year9 |    3.63963   .2736522    13.30   0.000     3.103205    4.176055
      year10 |   .3216964   .2745402     1.17   0.241    -.2164692     .859862
      year11 |  -2.921898   .2765643   -10.56   0.000    -3.464032   -2.379765
       _cons |    10.1434   .2828633    35.86   0.000     9.588915    10.69788
-------------+----------------------------------------------------------------
     sigma_u |  12.189694
     sigma_e |  5.7087583
         rho |  .82012265   (fraction of variance due to u_i)
------------------------------------------------------------------------------
F test that all u_i=0:     F( 76, 8496) =   528.82             Prob > F = 0.0000
```

*Tab. 15.*

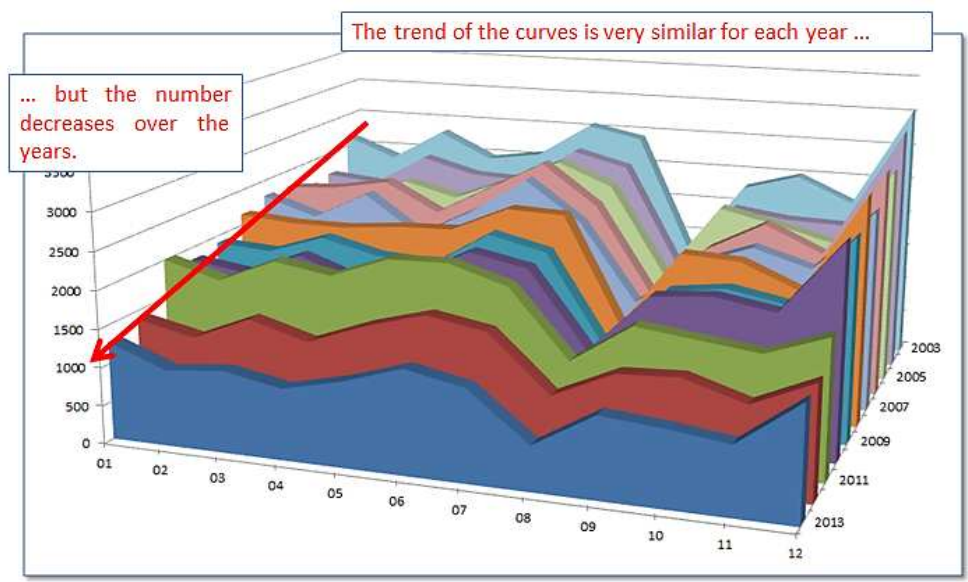

*Fig. 16.*

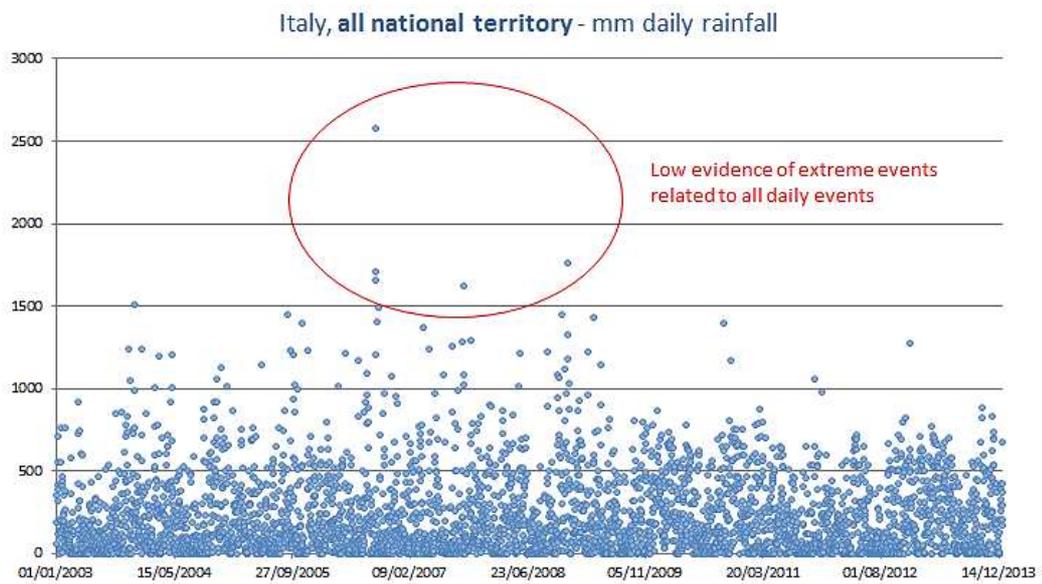

*Fig. 17.*

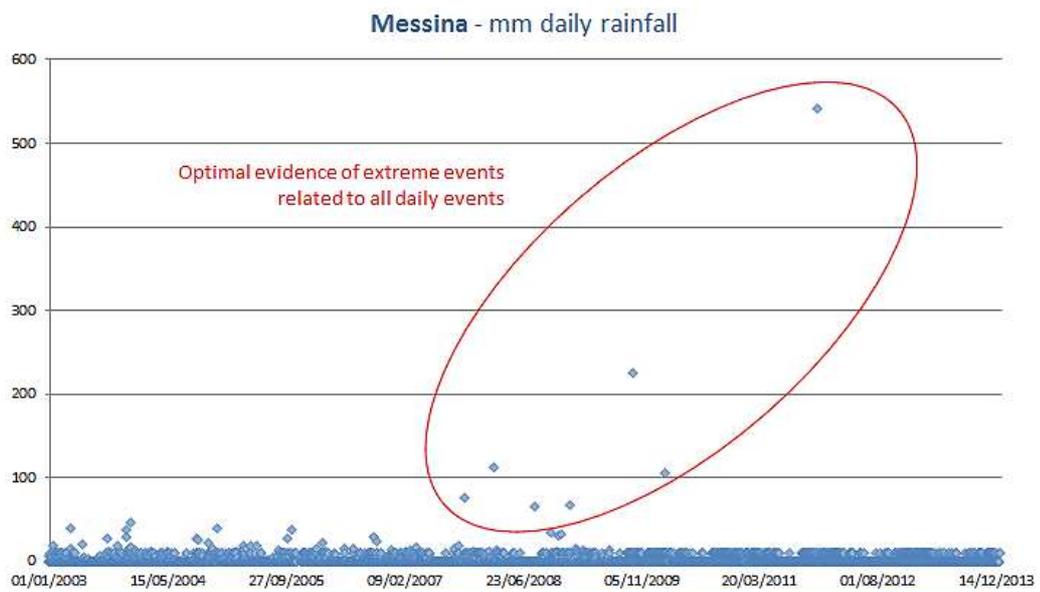

*Fig. 18.*

CONCLUSIONS.

A flood event causes in the months following an increase of appropriate policies to cover events?

Analysis of the data obtained by the regressions seems to tell us that there is no empirical evidence, due to the fact that the varibile that measures the presence of an event does not assume significance , while the control variables for months and years, independent by the presence of an extreme even, are significant.

We can deduce by the negative trend of issued policies that the increase of extreme events happening in recent years not only has an impact on the trend of the issue of insurance policies to cover these events, but even it seems to contract the market.

Where present, the insurance culture does not seem to suffer influence of the occurrence of floods in Italy, so the portion of the population that has so far been without insurance culture in relation to climate events continue to maintain this behavioral gap. On the other hand, even in geographic areas with greater attention to the protection of the risk does not seem to be noted significant increases in proportion to the increase of events.

THE PROPOSAL OF ANIA (National Insurance Enterprises Association).

1. As suggested by the OECD, governments should create the conditions so that private or public insurance instruments (or mixed) for natural disaster are made available in order to plan the economic resources to cover damages.
2. As suggested by the OECD for the best efficiency, even in systems where the risk taker is a public fund (eg Spain, the USA limited to floods, Turkey), the existing private insurance infrastructure should be used for the distribution of contracts and for the evaluation and settlement of claims.
3. For the sustainability of the system, is required a sort of geographical dispersion of risk (avoiding the so-called anti-selection ie the concentration of insured property only in areas at highest risk) and the attainment of a certain critical mass to achieve economies of scale.
4. The insurance/reinsurance private sector can allocate a certain amount of capital for catastrophe risk in Italy. If the diffusion of coverage was vary large, It will be insufficient to cover the demand. State intervention is therefore needed as a co-reinsurer. Note that the systems adopted in countries with a high exposure to catastrophic risks (like Italy) provides for a reduction of compensation in proportion in the case of events that exceed the capacity of the system (see California, Japan, Belgium) .

5. As a matter of principle, premiums proportional to the risk encourage prevention measures (as recommended by the OECD that suggests contributions to needy categories). However, it may provide the maximum prices to make insurance coverage popular and widely accessible.

ANIA aims to spread the culture of insurance to form conscious citizens.

Two arguments at the core of a project:
- prevention, useful to decrease the probability that a negative event may occur;
- mutuality, which allows protection offer to those who find in condition to deal with adverse events.

"Io&irischi" addresses to italian schools to promote a greater awareness of risk and a culture of prevention and its management during the life, with an important goal: educate to risk for educate to future.

In 2008, the OECD has given importance to these issues in the publication *Recommendation on Good Practices for Enhanced Risk Awareness and Education on Insurance Issues*, defining objectives and best practices for years to come, such as the diffusion of educational pathways, starting with the compulsory education, addressed to increase awareness and responsibility towards the potential risks to which they may be exposed individuals, and especially in high schools, the deepening of the concept of risk and the basic mechanisms of insurance.

"Io&irischi"  is aligned to these recommendations that, through the project "Io&irischi teens", guides the high school students to the understanding of "basic" insurance, such as "pure risks" and "speculative risks", the law of large numbers, the calculation of policy premiums, allowance and excess, in addition to supplementary social security schemes offered by the insurance sector.

INSURANCE CULTURE.

"The damage caused by natural disasters have increased dramatically over the last thirty years, especially due to the increasing economic value, and the insurance industry is the leading candidate for the distribution and management of the risks faced by households and firms and for the damages payment."
This is what says Prof. Antonio Coviello , researcher at the Institute for research on tertiary activities of the CNR and professor of economics and management of insurance companies at the Second University of Naples, author of the interesting volume " Natural disasters and insurance coverage ."

According to this study :
- between 1963 and 2012 a number of 782 Italian cities have suffered flooding and landslides;
- 1,563 victims of floods and 5,192 victims of landslides (dead, injured and missing);
- 421 227 persons displaced and homeless;
- in 2011 in the world more than 366 billion dollars in economic damage (about 280 billion Euros);
- direct damages caused by natural disasters in emerging countries achieve an average of 2.9% of GDP each year (Department of Economic Research of Munich Re);
- In these cases, the mandatory insurance coverage is the most appropriate instrument to manage economic damage from natural disaster: adopted in many European countries, is also advocated by the OECD, which suggests the introduction of appropriate regulatory frameworks in this field, to allow economic actors to plan possible interventions for prevention and preparedness;
- The insurance has surely a first indirect effect of damages reduction, as premiums are an incentive for taking preventive measures and give the respective risk a price, as well as obviuosly to directly support the reconstruction in case of a catastrophe.

BIBLIOGRAPHY.

- IPCC, Intergovernmental Panel on Climate Change - Climate Change 2013, WG1 Working Group I, 5th Assessment Report;
- IPCC, Managing the Risks of Extreme Events and Disasters to Advance Climate Change Adaptation, 2012;
- C. Cacciamani, Eventi estremi, entità e ricorrenza, Ecoscienza, n. 5 - 2013;
- Ministero dell'Ambiente Italiano, Classificazione dei Comuni Italiani in base al Livello di attenzione per il Rischio Idrogeologico, Servizio per la difesa del territorio, 2000;
- R. Manzato, Lo stato dell'arte sull'assicurazione dei rischi catastrofali, ANIA, 2014;
- OECD Governments, OECD Recommendation on good practices for enhanced risk awareness and education on insurance issues, 2008;
- A. Coviello, Calamità naturali e coperture assicurative, Flaccovio Ed, 2013;
- C. Courbage, W. R. Stahel, The Geneva Reports Risk and Insurance Research, Extreme events and insurance: 2011 annus horribilis, n. 5 - 2012;
- C. Rosenzweig, A. Iglesius, X. B. Yang, P. R. Epstein, E. Chivian, Climate change and extreme weather events - Implications for food production, plant diseases, and pests, University of Nebraska – Lincoln, NASA Publications - National Aeronautics and Space Administration, 2001;
- S. Hallegatte, Accounting for Extreme Events in the Economic Assessment of Climate Change, Fondazione Eni Enrico Mattei, 2005;
- C. G. Moles, J. R. Banga, K. Keller, A. S. Lieber, Global optimization of climate control problems using evolutionary and stochastic algorithms, 2003.